\documentclass[twocolumn, trackchanges, twocolappendix]{aastex701}

\begin{document}

\title{Neural Enhancement of the Traditional Wang–Sheeley–Arge Solar Wind Relation}

\author[orcid=0000-0001-8265-6254,sname='Mayank']{Prateek Mayank}
\affiliation{Space Weather TREC, University of Colorado, Boulder, CO 80303, USA}
\affiliation{University Corporation for Atmospheric Research, CPAESS, Boulder, CO 80301, USA}
\email[show]{pmayank@ucar.edu}  

\author[orcid=0000-0002-7862-6383, sname='Camporeale']{Enrico Camporeale} 
\affiliation{Space Weather TREC, University of Colorado, Boulder, CO 80303, USA}
\affiliation{Queen Mary University of London, London E1 4NS, United Kingdom}
\email{} 

\author[orcid=0000-0001-9035-3245, sname='']{Arpit K. Shrivastav} 
\affiliation{Southwest Research Institute, Boulder, CO 80302, USA}
\email{} 

\author[orcid=0000-0002-4989-475X, sname='']{Thomas E. Berger} 
\affiliation{Space Weather TREC, University of Colorado, Boulder, CO 80303, USA}
\affiliation{National Center for Atmospheric Research, High Altitude Observatory, Boulder, CO, 80301, USA}
\email{} 

\author[orcid=0000-0001-9326-3448, sname='']{Charles N. Arge} 
\affiliation{Heliophysics Science Division, NASA Goddard Space Flight Center, Code 671, Greenbelt, MD 20771, USA}
\email{} 

\begin{abstract}

The Wang--Sheeley--Arge (WSA) model has been the cornerstone of operational solar wind forecasting for nearly two decades, owing to its simplicity and physics-based formalism. However, its performance is strongly dependent on several empirical parameters that are typically fixed or tuned manually, limiting its adaptability across varying solar conditions.
In this study, we present a neural enhancement to the WSA framework (referred to as WSA{\texttt{+})} that systematically optimizes the empirical parameters of the WSA solar wind speed relation using in-situ observations within a differentiable physics-constrained pipeline.
The approach operates in two stages: first, a neural optimizer adjusts WSA parameters independently for each Carrington Rotation to better match the observed solar wind data. Then, a neural network learns to predict these optimized speed maps directly from magnetogram-derived features. This enables generalization of the optimization process and allows inference for new solar conditions without manual tuning.
WSA\texttt{+} preserves the interpretability of the original relation while significantly improving the match with OMNI in-situ data across multiple performance metrics, including correlation and error statistics. It consistently outperforms the traditional WSA relation across both low and high solar activity periods, with average improvements of approximately 40 percent. By integrating data-driven learning with physical constraints, WSA\texttt{+} offers a robust and adaptable enhancement, with immediate utility as a drop-in replacement in global heliospheric modeling pipelines.

\end{abstract}

\keywords{\uat{Solar wind}{1534}; \uat{Heliosphere}{711}; \uat{Space weather}{2037}; \uat{Solar coronal holes}{1484}}


\section{Introduction} \label{intro}

Space weather events driven by solar activity can severely impact satellites, aviation, communication systems, and electrical infrastructure on Earth. Among the various drivers, the solar wind is a continuous structured outflow of plasma from the Sun. It serves as a persistent background condition that modulates heliospheric dynamics and preconditions the near-Earth environment. It also strongly influences coronal mass ejection (CME) propagation \citep{winslow_2021_the, prateekmayank_2023_swasticme, kay2024} and CME-CME interactions \citep{lugaz_2017_the, Mayank2024, Smitha2025}. Accurate modeling of the ambient solar wind is therefore essential for both research and operational space weather forecasting.

A widely adopted tool for ambient solar wind modeling is the Wang--Sheeley--Arge \cite[WSA;][]{arge_2003_improved} model, which connects synoptic photospheric magnetic field observations to solar wind speed near the Sun via a potential field extrapolation and empirical relations. The WSA model forms the backbone of many forecasting pipelines, including WSA-ENLIL, which is used operationally by NOAA’s Space Weather Prediction Center \citep{2017SpWea..15.1383S} and the UK Met Office  \citep{2018cosp...42E.347B}. Its efficiency, simplicity, and empirical grounding have contributed to its continued relevance across multiple solar cycles.

However, the empirical nature of WSA also presents key limitations. The model’s wind speed prediction relies on a set of tunable parameters, which are often manually adjusted to optimize performance for individual Carrington Rotations (CRs). In practice, these parameters vary significantly across CRs and solar conditions \citep{issan_2023_bayesian}. Operational pipelines often adopt fixed average values, which limits adaptability and accuracy. Consequently, while per-CR parameter fitting can improve forecast accuracy, that approach remains incompatible with real-time forecasting workflows.

Several efforts have been made to improve the WSA model by refining its empirical formulation, adjusting boundary conditions, or addressing parameter uncertainty. Some studies have also focused on calibrating the speed relation. For instance, \cite{mcgregor_2011_the} proposed a revised formula based on Helios data. While, \cite{elliott_2022_improving} introduced empirical corrections using residual error maps tied to magnetic topology. Additionally, \cite{Riley2015} re-examined the expansion-factor term in WSA, showing that its influence can be limited under certain conditions, thereby motivating refinements to the empirical speed relation.

\cite{kumar_2025_on} and \cite{majumdar_2025_what} showed that varying the source surface height and using higher-quality magnetograms improves forecast accuracy. \cite{reiss_2020_forecasting} developed an adaptive system that integrates in-situ observations to better constrain near-Sun conditions. Together, these studies reveal the importance of flexibility and tuning. However, most approaches rely on manual calibration or static assumptions. This highlights the need for a generalizable, data-driven enhancement of WSA that can adapt across solar conditions in an automated and interpretable way.

In this work, we introduce a hybrid machine learning (ML) framework that overcomes this constraint by first optimizing WSA parameters per CR using a neural optimizer, and then training a neural network to generalize across CRs. The surrogate model predicts full 2D solar wind speed maps at 21.5 solar radius (0.1 AU), capturing structural variability while preserving physical interpretability. This hybrid ML approach allows us to address the limitation of tunable WSA parameters through learning the generalized pattern of their optimized set of values across the solar cycle. 

The structure of this paper is as follows: in Section \ref{sec:methodology}, we describe the two-stage modeling pipeline, including the neural optimizer and the neural network. In Section \ref{sec:results}, we present performance results across 129 CRs and analyze spatial and time series validation. Section \ref{sec:discussion} discusses implications for operational forecasting and limitations of the current approach. We conclude in Section \ref{sec:conclusion} with a summary and outlook.

\section{Methodology} \label{sec:methodology}

The WSA model computes solar wind speed by relating it to open fieldlines derived from observed synoptic magnetograms. This mapping relies on the Potential Field Source Surface (PFSS) model, which assumes a current-free inner-corona and solves the elliptic Laplace equation to compute the scalar potential of the magnetic field \citep{Altschuler1969}. From this solution, the flux-tube expansion factor at the source surface ($f_s$) and the minimum angular distance ($D$) to the coronal hole boundary are derived, both of which serve as key inputs to the empirical WSA wind speed relation:

\begin{equation}\label{wsa_eq}
    V_{wsa} = V_{min}+\frac{V_{max}}{(1+f_s)^{\alpha}} \times\,\bigg(1 - a1\,exp\, \bigg(-(D/w)^{\beta}\,\bigg)\bigg)^{a2}
\end{equation}

This empirical formula expresses the predicted WSA solar wind speed ($V_{wsa}$) as a function of $f_s$  and $D$, modulated by a set of tunable parameters: $V_{min}$, $V_{max}$, $\alpha$, $\beta$, $w$, $a1$ and $a2$. These parameters collectively determine the sensitivity of the model to changes in coronal magnetic topology, and effectively control how strongly the solar wind speed responds to changes in $f_s$ and $D$. While fixed default parameter values are often used in operational settings, such static configurations do not adapt to varying solar conditions. In this study, we adopt a standard default set: $V_{min} =250$, $V_{max}=750$, $\alpha=0.222$, $\beta=1.25$, $w=0.028$, $a1=0.8$ and $a2=3$, consistent with values used in previous studies \citep[e.g.,][]{mcgregor_2011_the, reiss_2019} and several contemporary solar wind modeling \citep[e.g.,][]{jenspomoell_2018_euhforia, mayank_2022_swastisw}.

\begin{figure*}
    \centering
    \includegraphics[width=\textwidth]{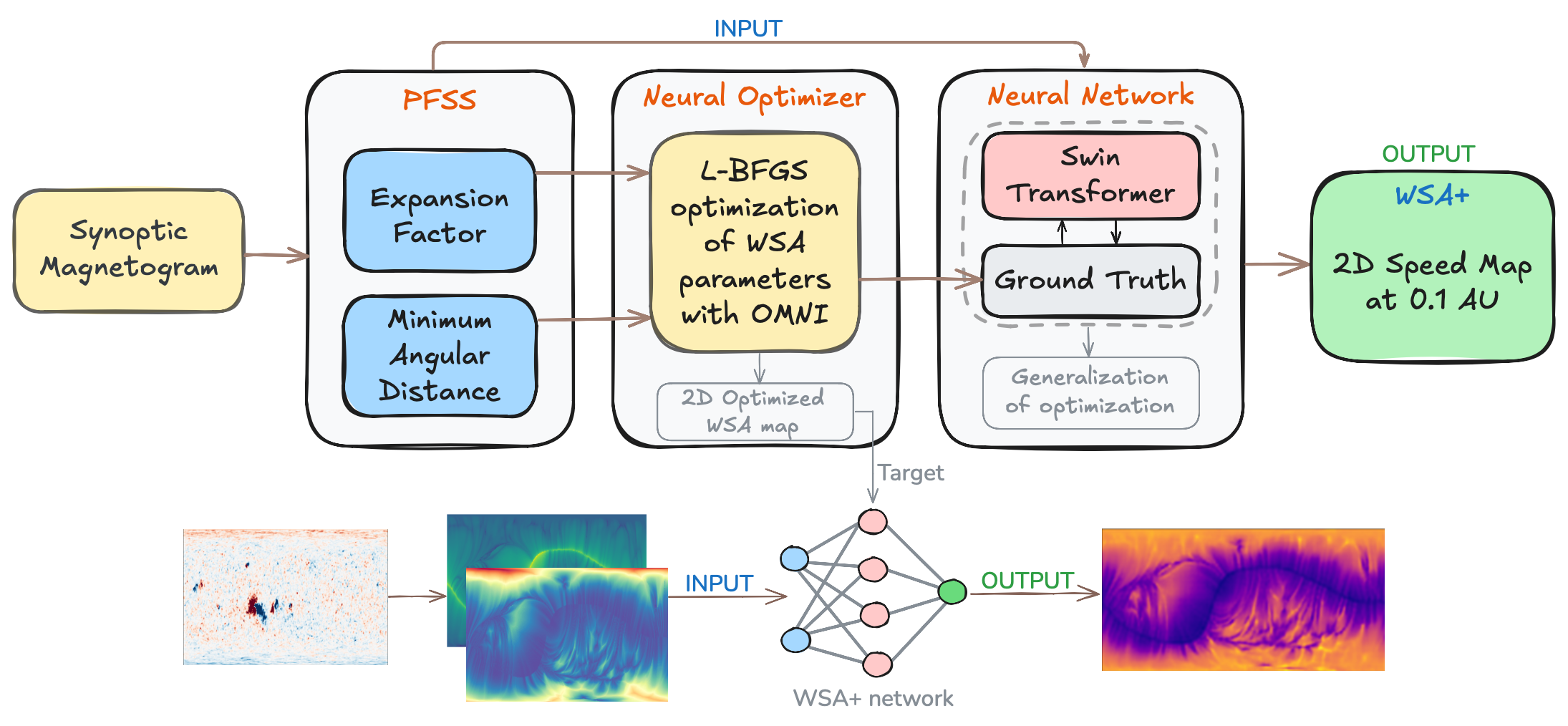}
    \caption{Schematic diagram of the undertaken method to optimize and generalize the WSA solar wind speed maps. PFSS is used to derive \textit{expansion factor} and \textit{minimum angular distance} maps from synoptic magnetograms. These 2D maps serve as input for both neural optimizer and neural network. L-BFGS scheme is employed to generate optimized 2D WSA maps by fitting the WSA parameters to OMNI data. Swin Transformer based neural network is trained on these optimized WSA maps to learn how to produce them directly from the PFSS-derived input maps.}
    \label{fig:flowchart}
\end{figure*}

\subsection{Neural Optimizer}
To address the limitations of fixed empirical parameters, we develop a differentiable optimization framework that couples the WSA speed relation with the Heliospheric Upwind Extrapolation \cite[HUX;][]{riley_2011_hux} model to produce wind speed at 1 AU. The empirical parameters in the WSA relation are treated as free variables constrained within physically plausible bounds around their default values (see Appendix \ref{Appendix1}). These parameters are optimized on a per–CR basis by minimizing a composite loss. This loss is computed from a set of statistical metrics that quantify the mismatch between predicted and observed solar wind speeds at L1, including root mean square error (RMSE), mean absolute error (MAE), dynamic time warping (DTW), and Pearson correlation coefficient (PCC).

This loss is minimized using the Limited-memory Broyden--Fletcher--Goldfarb--Shanno (L--BFGS) algorithm \citep{liu_1989_on}, a quasi-Newton method that approximates the inverse Hessian matrix to efficiently navigate the parameter space. Its low memory footprint and fast convergence make it particularly well-suited for optimizing low-dimensional problems with smooth, differentiable objective functions---like our case, where the number of empirical WSA parameters is small and the loss surface is well-behaved.

The resulting set of per-CR best-fit parameters is used to generate the optimized 2D WSA speed maps. These maps act as the ground truth for the neural network.

\subsection{Neural Network}
While per-CR optimization significantly improves the alignment between predicted and observed solar wind speeds, it requires in-situ data and is therefore not directly usable for forecasting. To generalize this optimization step and enable predictive capability, we train a neural network to infer the optimized WSA maps directly from magnetogram based PFSS--derived features. Figure \ref{fig:flowchart} demonstrates all the critical steps for the development of the WSA\texttt{+} model.

The input to the model consists of 2D maps of the expansion factor ($f_s$) and minimum angular distance ($D$), while the targets are the optimized WSA maps obtained from the neural optimization stage. Thus, the model is trained solely on the optimized 2D WSA maps --- not on 1D OMNI time series data or the best-fit numerical values of the seven WSA parameters.

We adopt a Swin Transformer \citep{Liu_2021_ICCV} architecture for this task, leveraging its ability to capture multi-scale spatial correlations and geometric context from the input maps. By learning from the 14 years of broad distribution of optimized maps, the network generates more realistic and smoothly varying output maps, in contrast to the parametrically optimized WSA maps, which are often overfitted to sub-Earth latitudes. This neural approach improves both the fidelity and generalizability of the solar wind speed maps derived from the WSA formulation. More information on the implemented Swin Transformer is in Appendix \ref{Appendix2}.

\subsection{Dataset and Training}
 In-situ solar wind speed observations from the 1-hour OMNI dataset \citep{Papitashvili_King_2020} are used for computing the optimization loss, covering a wide range of solar conditions from 2006 to 2024. All CR-based OMNI datasets are interpolated from 1-hour to 360 evenly spaced Carrington longitude points to match the resolution of the zero-point corrected synoptic GONG magnetograms. A 9-hour moving median filter is applied to suppress outliers and fill minor data gaps. CRs exhibiting large data gaps are excluded from the training set. Additionally, CRs impacted by halo coronal mass ejections (CMEs) are also excluded. 

The total dataset spans 129 CRs, among which 85 are used for training, 22 for validation, and 22 for testing. The training routine is implemented in PyTorch with reproducible random seed initialization. Optimization is carried out using the AdamW  optimizer \citep{Loshchilov_adamw}, and a cosine annealing scheduler is used to progressively reduce the learning rate. The training loss is computed from the spatial 2D map difference between model-predicted and CR-optimized WSA maps.

\section{Results} \label{sec:results}
The performance of WSA\texttt{+} is assessed through both spatial and temporal comparisons against the default WSA configuration and its CR-specific optimized variant. The following subsections demonstrate the results for the global solar wind speed structure through 2D maps, followed by in-situ time-series validation at the L1 point.

\subsection{2D Map Analysis}

\begin{figure*}
    \centering
    \includegraphics[width=\textwidth]{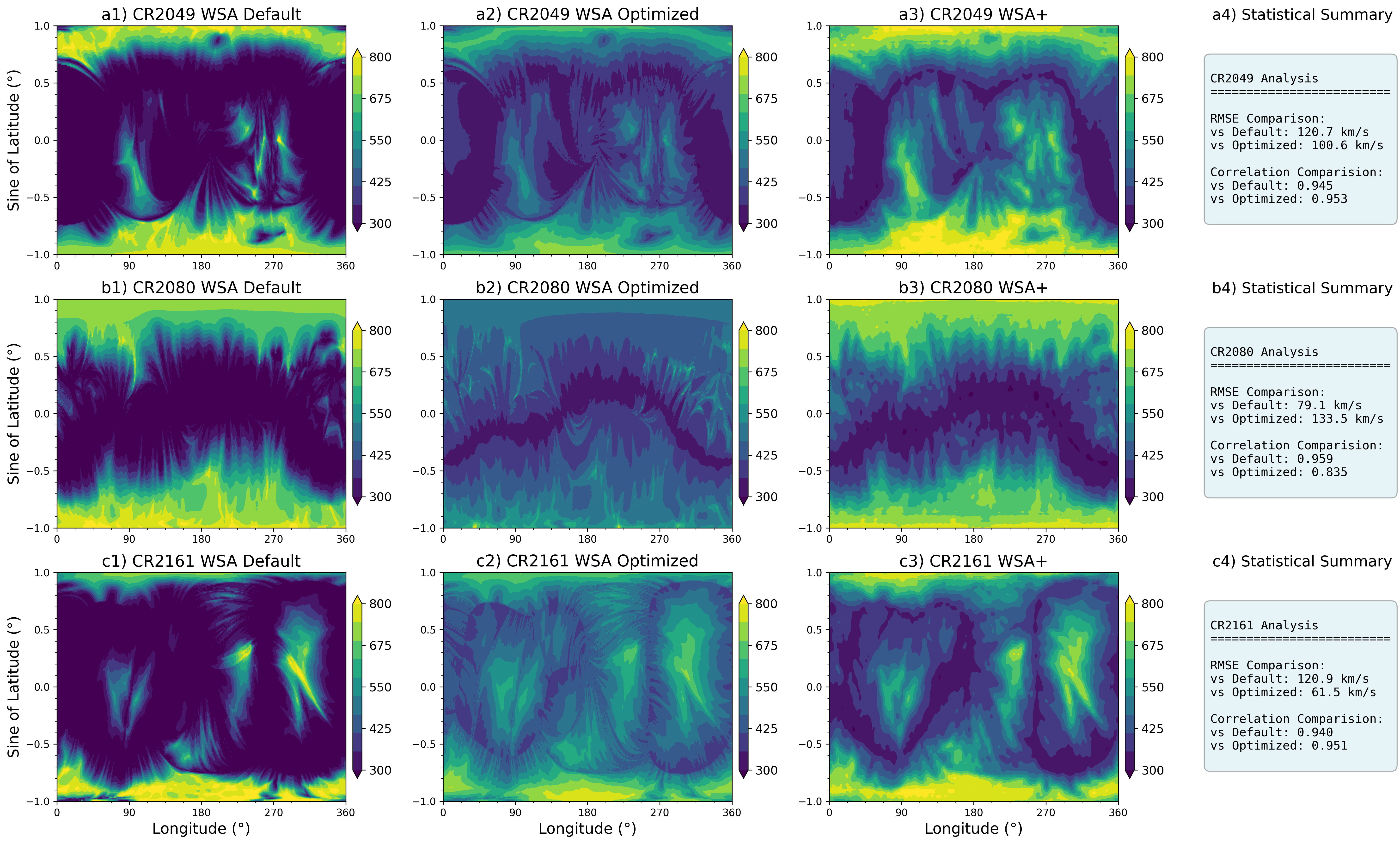}
    \caption{Comparison of Carrington maps of solar wind speed at 0.1 AU for three representative CRs: CR2049, CR2080, and CR2161, at different phases of the solar cycle. Each row corresponds to a different CR, while columns show outputs from default WSA (left), per-CR optimized WSA (middle), and WSA\texttt{+} (right).}
    \label{fig:2D_maps}
\end{figure*}

Comparing the global 2D solar wind maps offers an effective means to assess how well the modeled wind speed represents expected global coronal structures. Figure 2 shows solar wind speed maps for three CRs, corresponding to solar maximum, solar minimum, and an intermediate condition. Each row represents a different CR, with columns showing the outputs from the default WSA, the CR-specific optimized WSA, and the generalized WSA\texttt{+} model, respectively.

Despite differences in solar activity, the global structure of the wind maps across all three CRs retains consistent large-scale features. In particular, the complex equatorial patterns associated with the heliospheric current sheet are similarly captured in all cases by WSA\texttt{+}, suggesting that the model has learned the underlying topological relationship between expansion factor, minimum angular distance, and resulting solar wind speed. Although these maps appear similar at a broad scale, closer inspection reveals important distinctions.

A key difference emerges in the polar wind speed patterns during solar minimum (middle row). The CR-specific optimization, which adjusts WSA parameters solely based on 1D sub-Earth comparisons, fails to maintain physically realistic high-speed outflows at higher latitudes. Since this process focuses only on matching observations along the sub-Earth trajectory, it can unphysically suppress polar wind speeds. In contrast, WSA\texttt{+} captures and preserves the expected latitudinal gradient, with faster flows at higher latitudes. For this CR, the WSA\texttt{+} map is more correlated with the default (PCC = 0.96, RMSE = 79.1) than with the overfitted optimized output (PCC = 0.84, RMSE = 133.5).

Additionally, the default WSA maps exhibit a notable suppression of structural variation across the equatorial region, leading to overly smooth and slow outflow profiles. However, WSA\texttt{+} strikes a balance: retaining sharper transitions between slow and fast streams, restoring latitudinal gradients, and avoiding overfitting. This reflects its capacity to generalize across CRs and reproduce globally consistent and physically meaningful solar wind speed distributions. This intermediate positioning between the low-variance default and the overfitted optimized model highlights the robustness of WSA\texttt{+} in capturing both equatorial and high-latitude features under diverse solar conditions.

\subsection{In-situ Analysis}

\begin{figure*}
    \centering
    \includegraphics[width=0.8\textwidth]{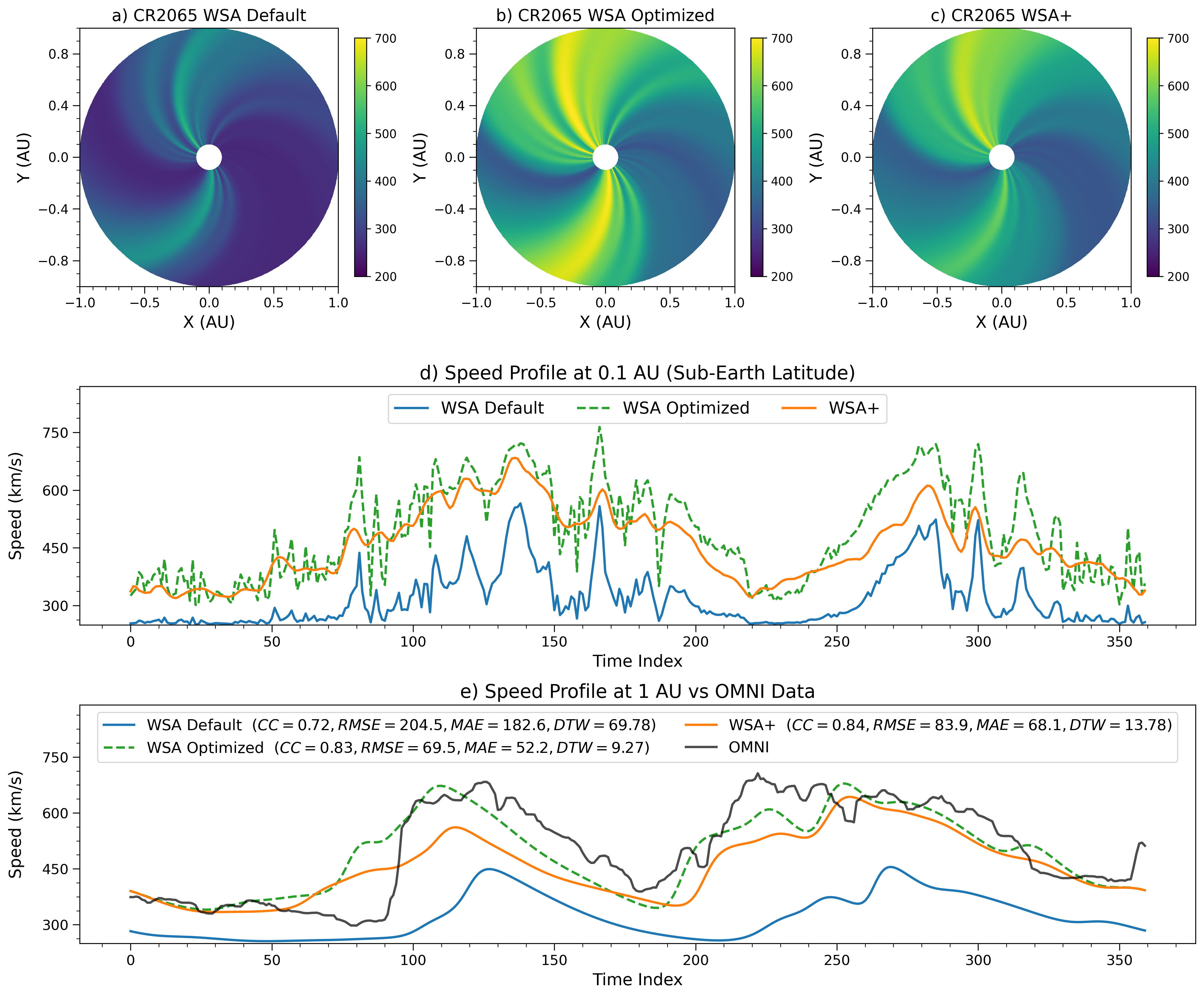}
    \caption{Comparison of solar wind speed profiles at sub-Earth latitudes for CR2065. All WSA variants are propagated from 0.1 AU to 1 AU using the HUX model. Panels (a–c) show wind speed maps along the elliptic plane, while panels (d) and (e) present the corresponding in-situ speeds at 0.1 AU and 1 AU, respectively.}
    \label{fig:in-situ_plots}
\end{figure*}

To complement the global structural assessment, we evaluate the model performance along the Sun–Earth line by comparing modeled solar wind speed with in-situ observations at L1. Figure 3 shows this comparison for a representative CR from the testing dataset, using three views: a meridional slice along the Sun–Earth spiral plane (top panel), and the corresponding in-situ speed profile at 0.1 AU (middle panel) and L1 (bottom panel). The color maps in the upper panel display the spatial evolution of the solar wind as computed by the default WSA, the CR-specific optimized WSA, and the generalized WSA\texttt{+}. The bottom panel shows the three curves representing predictions from each of these models overlaid with OMNI observations as the reference.

For the shown CR2065, the 2D speed profiles (top panel) from all three WSA variants exhibit similar structural morphology and relative stream positioning. However, notable differences emerge in the absolute wind speeds, particularly in the amplitude of high-speed streams. These differences are reflected in the in-situ speed profiles at 0.1 AU (middle panel), where the default WSA prediction (blue solid line) systematically underestimates the speed compared to the optimized variant (green dashed line). Although it more closely tracks the optimized solution, its shift is nonlinear, because of the influence of generalization across all CRs.

At L1 (bottom panel), the divergence among the models becomes more pronounced when compared against the OMNI observations. As expected, the CR-specific optimized WSA (PCC=0.83, RMSE=69.5, DTW=9.3) aligns closely with the observations. In contrast, the default WSA prediction (PCC=0.72, RMSE=204.5, DTW=69.8) continues to underestimate both the baseline and peak wind speeds, missing the timing and amplitude of high-speed stream arrivals by a larger margin. The WSA\texttt{+} model captures the baseline and local minima nearly as well as the optimized variant. However, the peak values fall short compared to the latter and the observations. Notably, it performs much better (PCC=0.84, RMSE=83.9, DTW=13.8) than the default WSA in estimating the minima, maxima, and arrival of the dominant high-speed streams, a critical factor in space weather forecasting.

\begin{figure*}
    \centering
    \includegraphics[width=\textwidth]{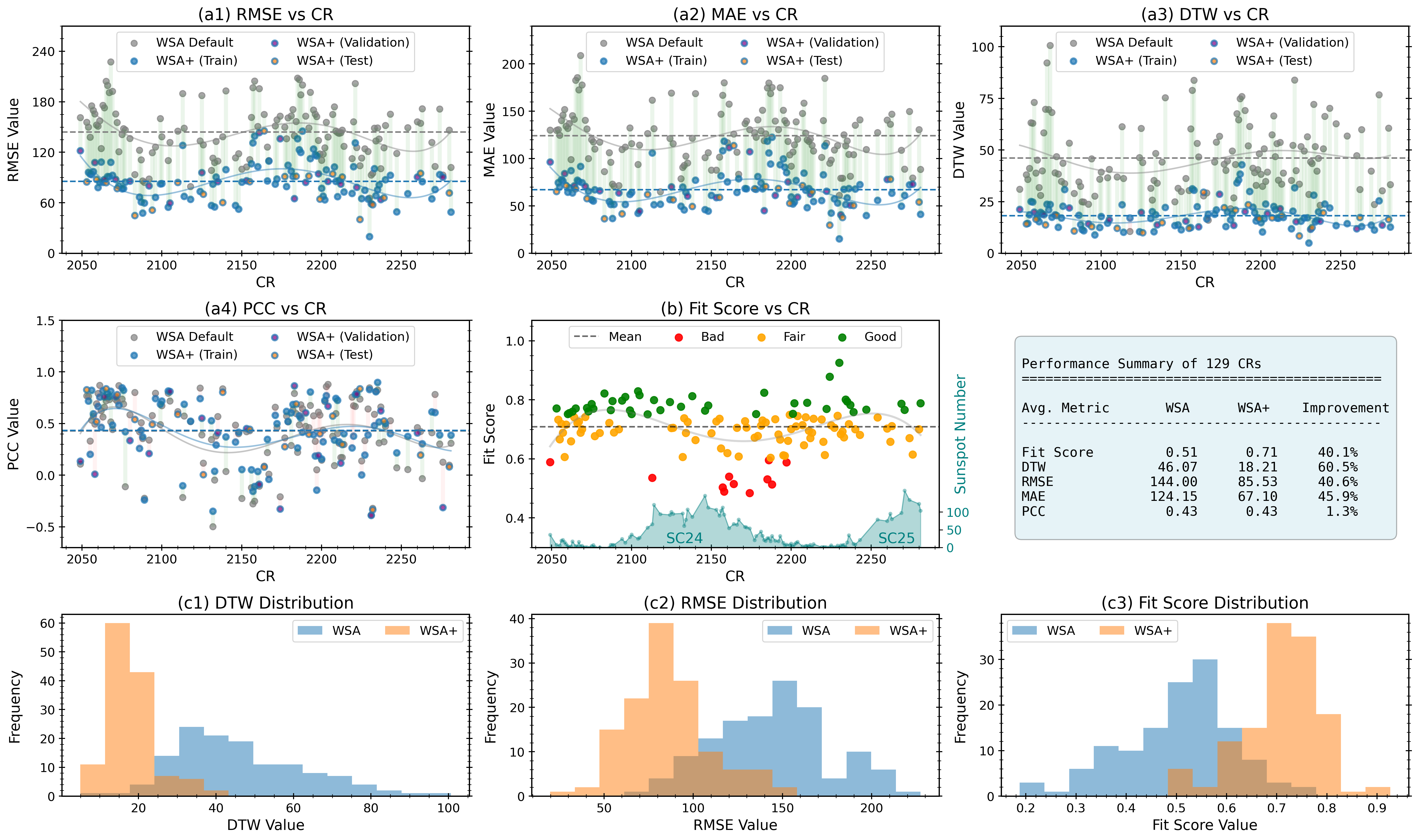}
    \caption{Comparison of model performance across 129 Carrington Rotations. Panels (a1-a4) show CR-wise variations of RMSE, MAE, DTW, and PCC for WSA and WSA\texttt{+}, with respect to OMNI data. Panel (b) plots the FIT score of WSA\texttt{+} with sunspot number and performance tier. Panels (c1-c3) show distribution histograms of DTW, RMSE, and FIT score.}
    \label{fig:metrics}
\end{figure*}

Consistently across CRs, WSA\texttt{+} demonstrates intermediate performance between the default and CR-specific optimized profiles. Figure 4 summarizes the comparison metrics between the default WSA and WSA\texttt{+} outputs against in-situ observations at L1, across considered CRs. The train, validation, and test datasets are shown in different colors. For the entire dataset, WSA\texttt{+} shows substantial improvements over the default WSA: a 60.5\% reduction in DTW, 45.9\% in MAE, and 40.6\% in RMSE, along with a marginal increase of 1.3\% in PCC. This pattern of significant gains in error-based metrics, with minimal change in linear correlation, suggests that WSA\texttt{+} effectively learns the underlying empirical relationship with $f_s$ and $D$, while preserving the structural integrity of the WSA formulation.

\section{Discussion} \label{sec:discussion}

\subsection{Solar Cycle Variability}
The performance of both default WSA and WSA\texttt{+}, as measured by the four evaluation metrics exhibits systematic variability across CRs. As shown in Figure 5(a1–a4), the error-based metrics (RMSE, MAE, and DTW) consistently show a substantial improvement for WSA\texttt{+}, while the PCC remains largely similar between the two models. However, all these metrics oscillate over time for both models, reflecting a quasi-periodic pattern. This consistent up-and-down trend across all four metrics indicates temporal sensitivity in model performance.


Each of these metrics captures a distinct aspect of fit quality, and none serves as a comprehensive indicator on its own. To address this, we define a composite FIT score, which averages the normalized values of all four metrics into a single 0 to 1 performance measure, where higher scores indicate better agreement with observations. Figure 5(b) illustrates the evolution of this score with solar activity. Interestingly, WSA\texttt{+} performs better during the ascending phases of the solar cycle, while showing reduced accuracy during the declining phases. Although the available CR span does not support a statistically robust conclusion, this behavior highlights the need for further investigation into solar cycle-dependent model performance, possibly linked to shifts in the global coronal magnetic topology.

Overall, WSA\texttt{+} outperforms the default WSA across the solar cycle, as evident from the metric distributions in Figure 5(c1 to c3). Moreover, the spread in metric values is consistently narrower for WSA\texttt{+}, reflecting greater robustness and stability in its predictive performance.

\subsection{Limitations and Follow Ups}
Despite the overall improvement achieved by WSA\texttt{+}, few challenges remain, which can further increase performance. The model’s output maps exhibit comparatively smoother features than the optimized WSA maps (see Figure 3(d)). This is because of the discrete patch-based learning of the Swin Transformer, which can inherently struggle to form very sharp global features. Additionally, the propagation of solar wind speed from 0.1 AU to 1 AU in this study relies on the HUX-based ballistic model, which introduces its own diffusive effects. These may further blur sharp features in the final time series.

Another limitation lies in the absence of a universally accepted evaluation metric for solar wind prediction. The reliance on multiple imperfect metrics introduces ambiguity in defining a true “best-fit”, which can mislead both model training and interpretation. Developing a physically informed and operationally relevant evaluation metric remains an important open challenge.

Looking forward, extensions involving spectral neural operators \citep{fanaskov_sno} may provide more physically compatible alternatives by learning mappings in continuous function space, rather than relying on discrete local patches.

\section{Conclusion} \label{sec:conclusion}
In this work, we demonstrated a unique two-stage method to enhance the WSA empirical relation through deep learning. By first optimizing the empirical coefficients on a per–CR basis and then generalizing across CRs using a Swin Transformer architecture, our approach captures physically meaningful relationships between magnetic topology and solar wind speed. This hybrid design allows the model to retain the interpretability of the original WSA formulation while improving adaptability.

The resulting model, WSA\texttt{+}, shows an overall improvement of 40\% across the entire dataset (39\% on the held-out test set) over the traditional WSA model. It serves as a direct, physically consistent enhancement to the WSA pipeline, and is made openly available through a Python package: wsaplus (\url{https://pypi.org/project/wsaplus/}). Designed for operational usability, WSA\texttt{+} can be seamlessly integrated into existing space weather modeling frameworks as a drop-in replacement for default WSA outputs, without requiring additional inputs or tuning.

Beyond its practical utility, WSA\texttt{+} demonstrates the potential of combining empirical domain knowledge with modern machine learning architectures to build transparent and reusable forecasting tools. Its performance stability and physically grounded structure make it a compelling step towards more accurate and interpretable space weather modeling.

\begin{acknowledgments}
This research was supported by the NASA Living with a Star Jack Eddy Postdoctoral Fellowship Program, administered by UCAR’s Cooperative Programs for the Advancement of Earth System Science (CPAESS) under award \#80NSSC22M0097.  This work was partially supported by NASA under awards No 80NSSC23M0192, 80NSSC20K1580, 80NSSC21K1555.
\end{acknowledgments}

\textit{Data Availablity Statement:} 
The used GONG synoptic magnetograms maps can be freely obtained from \url{https://gong.nso.edu/data/magmap/crmap.html}. The OMNI data are taken from the Goddard Space Flight Center, accessible at \url{https://spdf.gsfc.nasa.gov/pub/data/omni/}. The used CME catalog can be accessed from \url{https://wind.nasa.gov/ICME_catalog/ICME_catalog_viewer.php}.
The dataset presented in this paper, along with all the codes and the final model checkpoint, are archived at \url{https://doi.org/10.5281/zenodo.17309172}.

\vspace{2mm}

\textit{Code Sharing Statement:} All the developed codes, including training, testing and visualization, can be accessed through \url{https://github.com/PrateekMayank/wsaplus}.

\software{Pytorch \citep{NEURIPS2019_9015}, pfsspy \citep{Stansby2020}, pysdtw \citep{maghoumi2021deepnag}, timm \citep{Wightman_PyTorch_Image_Models}, dtw \citep{giorgino_2009_computing}.}

\appendix \label{sec:appendix}

This appendix provides technical details regarding the parameter optimization method and neural network architecture used in this study. Section A outlines the empirical parameter ranges and fitting strategy. Section B describes the WSA\texttt{+} model architecture based on Swin Transformer, along with training diagnostics.

\section{Optimization of Empirical Parameters} \label{Appendix1}
To generate CR-specific best-fit WSA maps, we employed a differentiable optimization framework that adjusts the seven tunable parameters in the empirical WSA formulation: $V_{min}$, $V_{max}$, $\alpha$, $\beta$, $w$, $a1$, and $a2$. For each CR, the HUX extrapolation model was used to propagate solar wind speeds from the WSA output at 0.1 AU to 1 AU, where the predicted time series was compared against OMNI observations. The optimization minimized a composite loss function based on RMSE, MAE, DTW, and PCC, using the L-BFGS method. Parameter values were constrained within physically plausible bounds, as listed in Table \ref{tab:parameter_ranges}.

\begin{table}[h]
\centering
\caption{Optimization bounds for empirical parameters in WSA equation.}
\begin{tabular}{lcc}
\hline
\textbf{Parameter} & \textbf{Description} & \textbf{Range} \\
\hline
$V_{\min}$ & Minimum wind speed & [250, 300] \\
$V_{\max}$ & Maximum wind speed & [600, 900] \\
$\alpha$ & Exponent on $f_s$ & [0.01, 0.5] \\
$\beta$ & Exponent on $D$ & [0.01, 3.0] \\
$w$ & Normalization factor for $D$ & [0.01, 0.05] \\
$a_1$ & Coefficient 1 for $D$ term & [0.7, 0.9] \\
$a_2$ & Exponent on $D$ term & [0.5, 6.5] \\
\hline
\end{tabular}
\label{tab:parameter_ranges}
\end{table}

\begin{figure}
    \centering
    \includegraphics[width=\columnwidth]{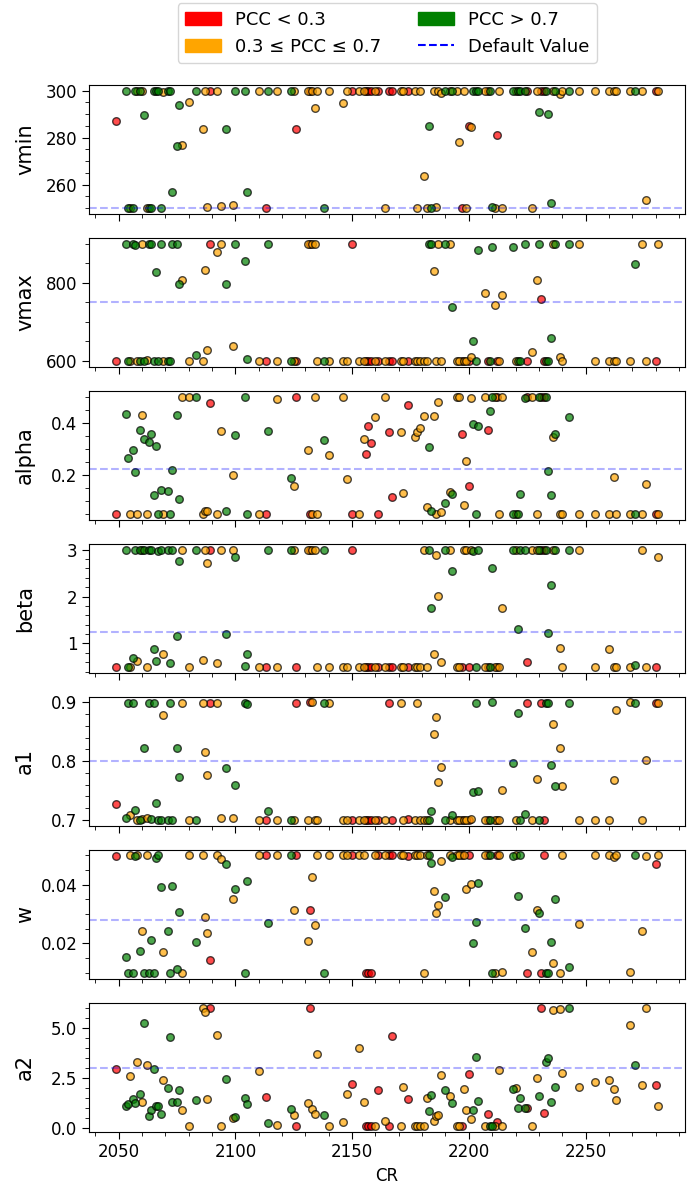}
    \caption{
    Distributions of the seven empirical WSA parameters optimized for each Carrington Rotation. The points are color-coded with their Pearson correlation coefficient (PCC) with in-situ observations.
    }
    \label{fig:params}
\end{figure}

\vspace{3mm}
\begin{figure}
    \centering
    \includegraphics[width=\columnwidth]{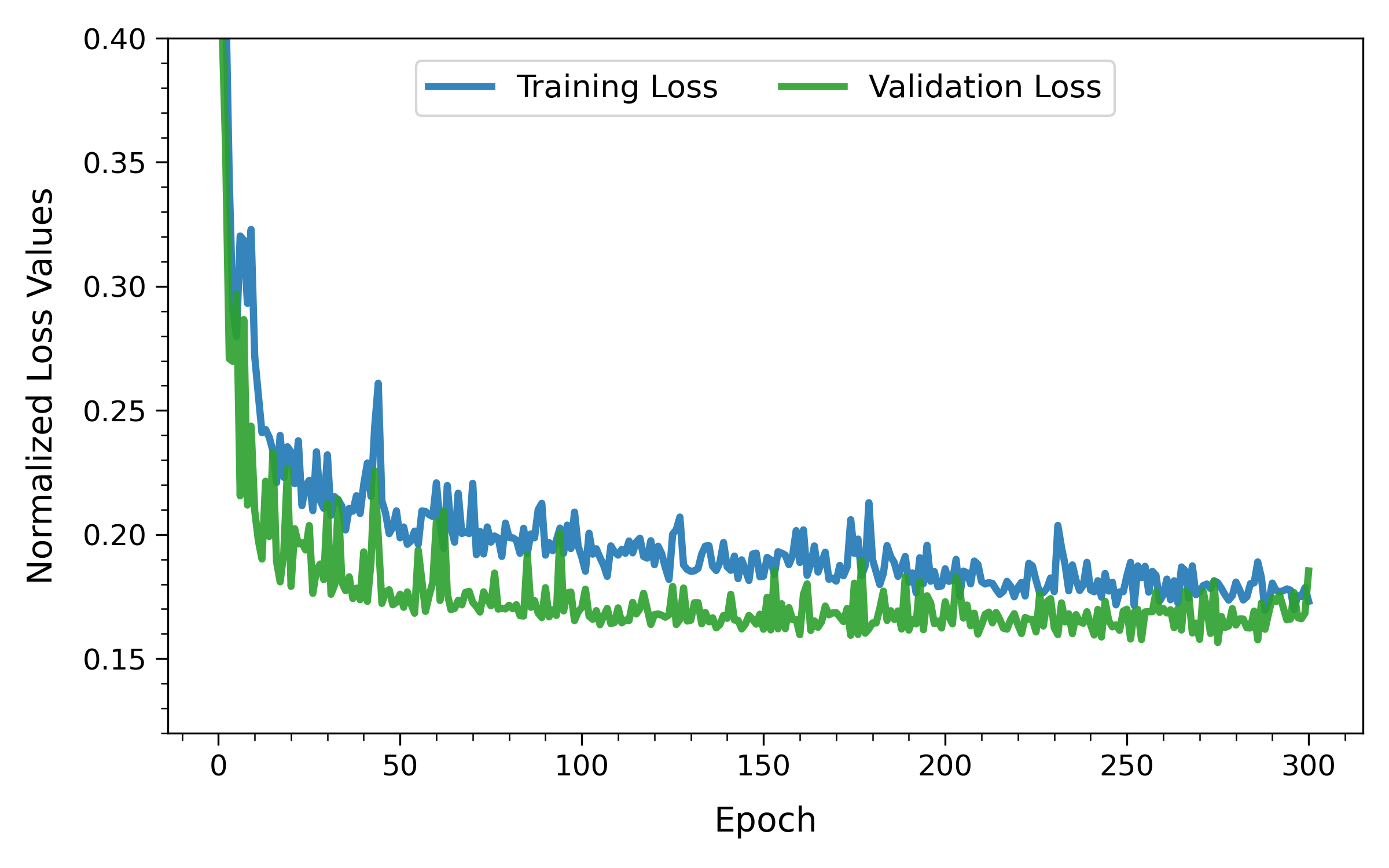}
    \caption{Evolution of training and validation losses over 300 epochs of WSA\texttt{+} model training.}
    \label{fig:loss_trend}
\end{figure}

The set of optimized values for the seven empirical parameters is shown in Figure \ref{fig:params}. Each point represents the parameter configuration that yielded the lowest loss between the WSA+HUX and observed solar wind speed at 1 AU for a given CR. The color of each point indicates the corresponding PCC value with OMNI observations. The distribution shows no clear pattern across CRs, suggesting a lack of systematic variation with solar cycle phase. Notably, several parameters frequently converge to their prescribed bounds. While one might infer that extending these ranges could improve performance, iterations indicate that values beyond the current limits do not yield better fits and often lead to unphysical outcomes. For instance, the minimum wind speed at 0.1 AU cannot reasonably exceed 500 km/s or drop far below 200 km/s.

This behavior, where best-fit parameter values frequently reach the bounds of their prescribed ranges, has also been observed in prior studies by \citet{riley_2015_on} and \citet{issan_2023_bayesian}. \citet{riley_2015_on} optimized five WSA parameters against ACE/WIND observations using the PCC metric, while \citet{issan_2023_bayesian} used Bayesian inference with Markov Chain Monte Carlo to estimate the distribution of five most influential WSA parameters. These findings, consistent with our results, likely reflect the limited identifiability of WSA parameters when tuned using 1D in-situ time series data within physically constrained bounds.

\section{Model Architecture and Training} \label{Appendix2}
\begin{figure*}
    \centering
    \includegraphics[width=\textwidth]{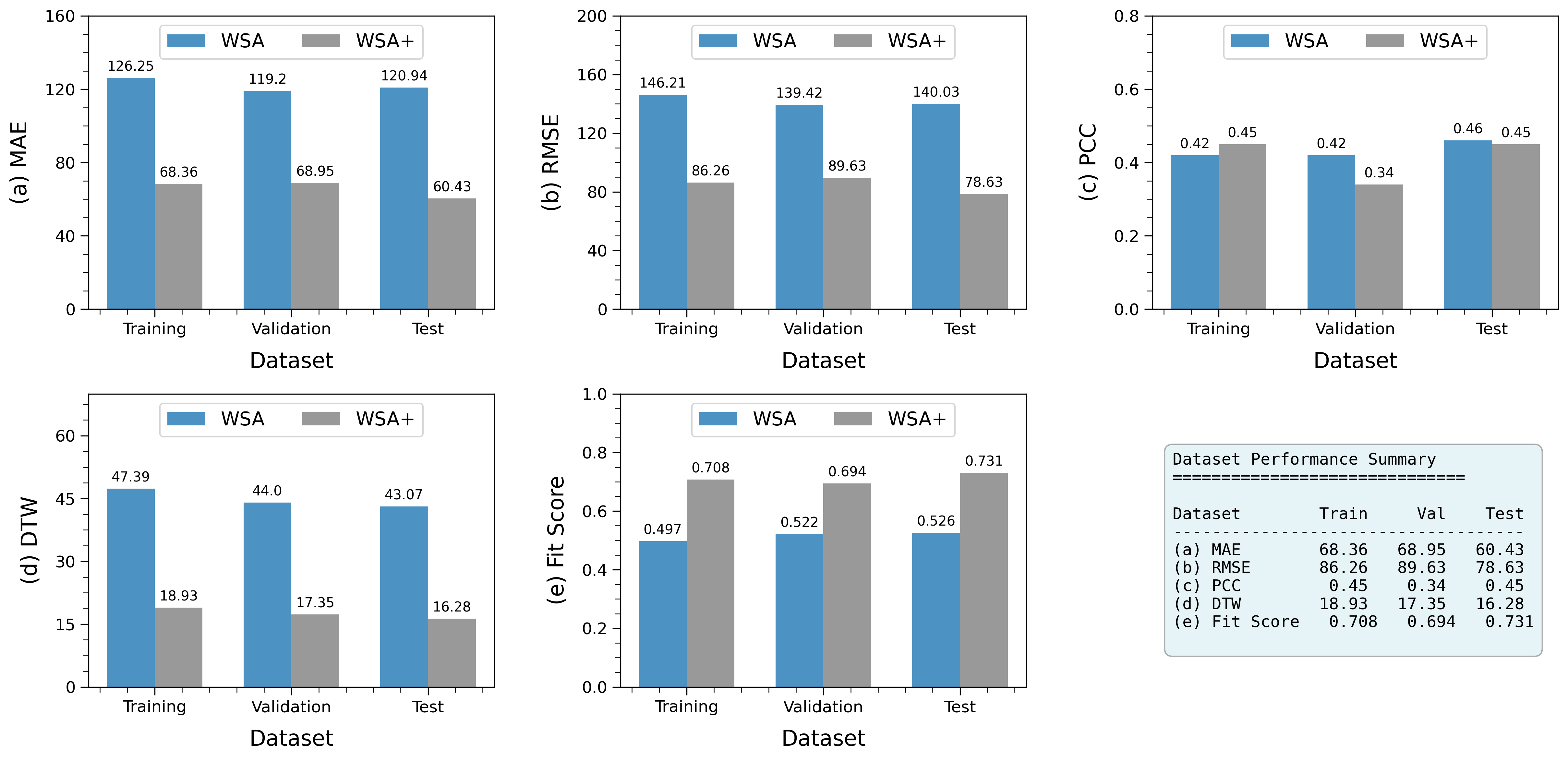}
    \caption{Metric-wise comparison of WSA\texttt{+} against default WSA across training, validation, and test sets. }
    \label{fig:dataset_comparison}
\end{figure*}

The WSA\texttt{+} model adopts a Swin Transformer based encoder–decoder architecture. The Swin Transformer is a hierarchical Vision Transformer designed to overcome domain-specific challenges to visual data, such as high spatial resolution and scale variation in features \citep{Liu_2021_ICCV}. It achieves computational efficiency by computing self-attention within non-overlapping local windows and enhances representational power through a shifted window mechanism that allows interaction across neighboring windows. The hierarchical nature of the architecture enables multi-scale feature extraction while maintaining linear complexity with image size. For the WSA\texttt{+} implementation, the model receives a $360 \times 180 \times 2$ input tensor composed of the PFSS-derived expansion factor ($f_s$) and minimum angular distance ($D$) maps, and outputs a $360 \times 180$ tensor representing the WSA\texttt{+} output map.

The encoder consists of four hierarchical stages of Swin Transformer blocks from a pretrained \textsc{swin-small-patch4-window7-224} backbone, progressively downsampling the input while enriching feature representations. These stages capture multi-scale spatial features of the input coronal topology. The decoder mirrors this hierarchy using three learned upsampling stages implemented via transposed convolutions (\textsc{ConvTranspose2d}), each followed by skip connections that merge with encoder outputs through 3×3 circular-padded convolution layers. A custom \textsc{RefineHead}, composed of residual convolutional blocks, further sharpens the decoded features prior to final projection. The output is passed through a PReLU activation to preserve non-linearity. The model is trained using mean squared error loss between the predicted WSA\texttt{+} map and the CR-optimized WSA map as the ground truth.

The model was trained for 300 epochs using the Adam optimizer with a batch size of 8 and an initial learning rate of $10^{-5}$, progressively decayed using a cosine annealing scheduler. Figure \ref{fig:loss_trend} illustrates the training and validation loss evolution across epochs, showing a consistent downward trend with no sign of overfitting, confirming effective convergence.

Figure \ref{fig:dataset_comparison} compares the model performance across training, validation, and test sets using five evaluation metrics. WSA\texttt{+} consistently outperforms the default WSA--HUX pipeline in RMSE, MAE, and DTW, with improvements sustained across all dataset splits. While the improvement in PCC is not consistent, the overall FIT score is consistently higher for WSA\texttt{+} across all datasets, demonstrating reliable predictive performance. However, the ratio of metric values of WSA and WSA\texttt{+} is slightly different for training, validation and test sets. This variation arises primarily because the ratio depends on which CRs are included in each dataset. Since the fidelity of the GONG+PFSS+WSA+HUX pipeline varies from CR to CR, the relative performance metrics will also exhibit similar variation. A summary of these statistics is also provided within Figure \ref{fig:dataset_comparison}.

\bibliography{WSA+}
\bibliographystyle{aasjournalv7}



\end{document}